\documentstyle[12pt]{article}
\newcommand{\nc}{\newcommand}

\nc{\dbar}{\bar{\partial}}
\nc{\be}{\begin{equation}}
\nc{\ee}{\end{equation}}

\catcode`\@=11

\@addtoreset{equation}{section}

\def\@normalsize{\@setsize\normalsize{15pt}\xiipt\@xiipt
\abovedisplayskip 14pt plus3pt minus3pt%
\belowdisplayskip \abovedisplayskip
\abovedisplayshortskip  \z@ plus3pt%
\belowdisplayshortskip  7pt plus3.5pt minus0pt}
\def\small{\@setsize\small{13.6pt}\xipt\@xipt
\abovedisplayskip 13pt plus3pt minus3pt%
\belowdisplayskip \abovedisplayskip
\abovedisplayshortskip  \z@ plus3pt%
\belowdisplayshortskip  7pt plus3.5pt minus0pt
\def\@listi{\parsep 4.5pt plus 2pt minus 1pt
            \itemsep \parsep
            \topsep 9pt plus 3pt minus 3pt}}

\def\underline#1{\relax\ifmmode\@@underline#1\else
        $\@@underline{\hbox{#1}}$\relax\fi}
\@twosidetrue
\relax

\catcode`@=12

\catcode`\@=11

\def\section{\@startsection{section}{1}{\z@}{3.5ex plus 1ex minus
   .2ex}{2.3ex plus .2ex}{\large\bf}}

\def\ps@headings{\def\@oddfoot{}\def\@evenfoot{}
\def\@oddhead{\hbox{}\hfill
        \makebox[.5\textwidth]{\raggedright\ignorespaces --\thepage{}--
        \hfill }}
\def\@evenhead{\@oddhead}
\def\subsectionmark##1{\markboth{##1}{}}
}

\ps@headings

\catcode`\@=12

\relax

\def\figcap{\section*{Figure Captions\markboth
        {FIGURECAPTIONS}{FIGURECAPTIONS}}\list
        {Fig. \arabic{enumi}:\hfill}{\settowidth\labelwidth{Fig. 999:}
        \leftmargin\labelwidth
        \advance\leftmargin\labelsep\usecounter{enumi}}}
 \relax
\def\tablecap{\section*{Table Captions\markboth
        {TABLECAPTIONS}{TABLECAPTIONS}}\list
        {Table \arabic{enumi}:\hfill}{\settowidth\labelwidth{Table 999:}
        \leftmargin\labelwidth
        \advance\leftmargin\labelsep\usecounter{enumi}}}
 \relax
\def\reflist{\section*{References\markboth
        {REFLIST}{REFLIST}}\list
        {[\arabic{enumi}]\hfill}{\settowidth\labelwidth{[999]}
        \leftmargin\labelwidth
        \advance\leftmargin\labelsep\usecounter{enumi}}}
 \relax

\catcode`\@=11

\def\ps@headings{\def\@oddfoot{}\def\@evenfoot{}
\def\@oddhead{\hbox{}\hfill
        \makebox[.5\textwidth]{\raggedright\ignorespaces --\thepage{}--
        \hfill }}
\def\@evenhead{\@oddhead}
\def\subsectionmark##1{\markboth{##1}{}}
}

\ps@headings

\relax

\def\firstpage#1#2#3#4#5#6{
\begin{document}

\begin{titlepage}
\nopagebreak
\title{\begin{flushright}
       \vspace*{-1.8in}
       {\normalsize IC/96/114  \\[-9mm]SISSA-112/96/EP}\\[-9mm]
        {\normalsize hep-th/9607131}\\[4mm]
\end{flushright}
\vfill
{\large \bf #3}}
\author{\large #4 \\ #5}
\maketitle
\vskip -7mm
\nopagebreak
\begin{abstract}
{\noindent #6}
\end{abstract}
\vfill
\begin{flushleft}
\rule{16.1cm}{0.2mm}\\[-3mm]
%$^{\star}${\small Research supported in part by\vspace{-4mm}
%the National Science Foundation under grant
%PHY--93--06906,\newline in part by the EEC contracts \vspace{-4mm}
%SC1--CT92--0792 and CHRX-CT93-0340,
%and in part by CNRS--NSF
%grant INT--92--16146.}\\[-3mm]
%$^{\dagger}${\small Laboratoire Propre du CNRS UPR A.0014.}\\
%July 1995
$^{\star}${\small Permanent address: Institute of Mathematical Sciences,
Madras - 600 113, India.}
\end{flushleft}
\thispagestyle{empty}
\end{titlepage}}
\newcommand{\dal}{\raisebox{0.085cm}
{\fbox{\rule{0cm}{0.07cm}\,}}}
\newcommand{\dt}{\partial_{\langle T\rangle}}
\newcommand{\dtbar}{\partial_{\langle\bar{T}\rangle}}
\newcommand{\al}{\alpha^{\prime}}
\newcommand{\mst}{M_{\scriptscriptstyle \!S}}
\newcommand{\mpl}{M_{\scriptscriptstyle \!P}}
\newcommand{\dv}{\int{\rm d}^4x\sqrt{g}}
\newcommand{\lv}{\left\langle}
\newcommand{\rv}{\right\rangle}
\newcommand{\ph}{\varphi}
\newcommand{\sbar}{\,\bar{\! S}}
\newcommand{\xbar}{\,\bar{\! X}}
\newcommand{\fbar}{\,\bar{\! F}}
\newcommand{\zbar}{\,\bar{\! Z}}
\newcommand{\tbar}{\bar{T}}
\newcommand{\ubar}{\bar{U}}
\newcommand{\ybar}{\bar{Y}}
\newcommand{\phb}{\bar{\varphi}}
\newcommand{\cm}{Commun.\ Math.\ Phys.~}
\newcommand{\pr}{Phys.\ Rev.\ D~}
\newcommand{\prl}{Phys.\ Rev.\ Lett.~}
\newcommand{\pl}{Phys.\ Lett.\ B~}
\newcommand{\ibar}{\bar{\imath}}
\newcommand{\jbar}{\bar{\jmath}}
\newcommand{\np}{Nucl.\ Phys.\ B~}
\newcommand{\e}{{\rm e}}
\newcommand{\gsi}{\,\raisebox{-0.13cm}{$\stackrel{\textstyle
>}{\textstyle\sim}$}\,}
\newcommand{\lsi}{\,\raisebox{-0.13cm}{$\stackrel{\textstyle
<}{\textstyle\sim}$}\,}
\date{}
\firstpage{95/XX}{3122}
{\large\sc D-branes and the conifold singularity} {E. Gava$^{a,b}$, T. 
Jayaraman $^{c,\star}$,
K.S. Narain$^{ b}$ $\,$and$\,$
M.H. Sarmadi$^{\,b}$}%\\[-3mm]
%{\normalsize\sl
%$^a$Centre de Physique Th\'eorique, Ecole Polytechnique,$^\dagger$
%F-91128 Palaiseau, France\\[-3mm]
{\normalsize\sl
$^a$Instituto Nazionale di Fisica Nucleare, sez.\ di Trieste,
Italy\\[-3mm]
\normalsize\sl $^b$International Centre for Theoretical Physics,
I-34100 Trieste, Italy\\[-3mm]
\normalsize\sl $^c$SISSA, I-34100 Trieste, Italy\\[-3mm]}
%\normalsize\sl $^d$Department of Physics, Northeastern
%University, Boston, MA 02115, U.S.A.}
{We analyze in detail the description of type IIB theory on a 
Calabi-Yau three-fold near a conifold singularity in terms of 
intersecting D-branes. In particular we study 
the singularity structure of higher derivative $F$-terms of 
the form $F_g W^{2g}$ where $W$ is the gravitational superfield. This 
singularity is expected to be due to a one -loop contribution from a charged
soliton hypermultiplet becoming massless at the conifold point. In the 
intersecting D-brane description this soliton is described by an open 
string stretched between the two D-branes. After identifying the 
graviphoton vertex as a closed string operator we show that $F_g$'s have 
the expected singularity structure in the limit of vanishing soliton mass.}
\section{Introduction}
    Solitonic objects play an important role in the understanding of
dualities in field theory and string theory. In particular, in the case
of N = 2 supersymmetric 
type II - heterotic string duality in four dimensions, an important
role is played by the appearance of massless charged hypermultiplets of
the gauge fields of the theory , arising from the zero-mass limit of
solitonic objects in the theory. In the case of Calabi-Yau
compactification of type IIB theories, the solitons are associated with 
Ramond-Ramond
three-branes wrapped around the three-cycles of the Calabi-Yau, that shrink to 
zero size in the conifold limit\cite{strom}.
Following Polchinski\cite{pol} these Ramond-Ramond solitons can be described 
as Dirichlet 3-branes.
However it is difficult to use Polchinski's prescription for these 
curved D-branes for a concrete string theory computation, especially
where we would like to
think of these D-branes as massless states, appearing for instance in
the internal line of a corresponding field theory computation.
An important step was taken
by Bershadsky, Sadov and Vafa \cite{bsv} when they proposed that by
using T-duality these solitons could be described as solitonic strings, or
D-strings. These could then be transformed into fundamental strings using 
the SL(2,Z) duality of type IIB theory. In the end therefore the solitonic
states would be accessible as perturbative string states, where the
string propogates in a manifold dual to the original compactification,
referred to as the D-manifold. 
   Thus in this picture the effects of solitons interacting with other
states in the theory would be perturbatively accessible. 

  In this note we examine one such application of the proposal of 
\cite{bsv}. We compute the topological amplitudes
arising from the higher-derivative F-terms of the form
$F_{g}(Z)W^{2g}$, where $W$ is the $N=2$ gravitational superfield
and $F_g$ depends on chiral vector superfields $Z$ (upto a holomorphic 
anomaly).
These occur at genus $g$ respectively in the type II theory \cite{agnt1}.
While in general the existence of such terms was shown, it is extremely
difficult to compute exactly the 
coefficient $F_g$ at all orders.
However,
it was shown in \cite{agnt} that in the dual heterotic string theory
these amplitudes occur at one-loop (with the exeception of 
$F_0$ and $F_1$ which also receive tree-level contributions)
and obey the same holomorphic anomaly
equations as in the type II theory. Moreover the leading singularities
of $F_g$ were studied in the conifold limit, and it was shown that they
are universal poles of order $2g-2$ with coefficients that are given 
by the Euler number of the moduli space of genus-$g$ Riemann surfaces,
in line with the argument of \cite{deb} that the physics near the
conifold singularity is governed by the $c=1$ string theory at
the self-dual point.

  It was also pointed out in \cite{agnt} that from the effective field
theory point of view,
the divergence comes from a one-loop graph with the
massless charged hypermultiplet
going through the loop. Hence in the type II theory
it was expected that the leading singularities in $F_g$
should arise from a one-loop diagram involving the would-be massless
charged black-hole of the conifold singularity in the internal line.
The possibility of  such a computation is precisely what is provided by
the D-brane description of the neighbourhood of the conifold 
singularity. In particular,
the leading singularity in $F_g$ should arise from a one-loop
computation. However since
the open string sector is what describes the soliton background, the
one-loop refers to the open-string one loop, viz. the annulus. 
   We show in this note that we reproduce the leading singularity
behaviour of $F_g$'s as expected in the
conifold limit. 
 
  In the next section we briefly review
the D-manifold description of the type
II theory near the conifold singularity and explicitly display the 
graviphoton operators that lie in the same space-time supersymmetry 
multiplet as the graviton. In section $3$, we show the detailed
computation of $F_g$ and show that it agrees with the previously
calculated expressions in \cite{agnt}. 

\section{The graviphoton operator} 
  We shall describe here only the D-manifold necessary for examining the
region close to a single conifold singularity. We will not describe more
general possibilities which include conifold singularities and we refer
the interested reader to \cite{bsv} and \cite{ov}. The region near the
confold singularity is described by two five-branes that intersect each
other. The first we may take to be specified by Dirichlet boundary
conditions on the directions $x_6$ to $x_9$ given by
$$
(x_6, x_7, x_8, x_9 ) = v
$$ 
The rest of the directions will satisfy Neumann boundary conditions.
The second D-brane is given by specifying the boundary condition
$$
(x_4, x_5, x_8, x_9) = (0, 0, 0, 0)
$$
while all the other co-ordinates satisfy Neumann boundary conditions.
When $v=0$, it is clear that the two D-branes will intersect on 
a 3 + 1 world-volume. We can now examine the different sectors of the
string whose end-points lie on the two D-branes, specifying the
boundary conditions on both ends of the string. $x_8$ and $x_9$ are 
DD directions, while $x_4$, $x_5$, $x_6$, $x_7$ are ND directions
and DN directions. $x_0$ to $x_3$ obey NN boundary conditions.
The massless charged state of interest, which corresponds to  a
massless hypermultiplet comes from the ground state of the string
between these two D-branes in the limit of the parameter $v$ going
to $0$. The U(1) gauge symmetry under which this state is charged
arises from a linear combination of the two boundary U(1) gauge fields
that are associated to the two D-branes. 

As is well known, in each five-brane sector the
correct space-time supersymmetry operator is given by a linear combination of
the left and the right moving space-time supersymmetry generators 
of the form $Q + M{\tilde Q}$, where M is made up of a
product of gamma matrices.
Since we are considering intersecting five-branes we have to find the
common eigenvalues of the corresponding M-matrices\cite{pol1}.
This gives, as we shall see below,
the effective space-time supersymmetry in the intersecting 
world-volume to be N=2. 
We will use this operator to act on the graviton operator
and obtain the explicit expression for the anti-self-dual part 
of the gravi-photon vertex operator.

For the first five-brane the matrix M appearing in the supersymmetry
operator is: 
\begin{equation}
M^{(1)} = \prod\gamma^{6}\gamma^{7}\gamma^{8}\gamma^{9}\gamma^{11}
\end{equation}
Acting on a spinor in 10 dimensions, this matrix has two $+1$ and two
$-1$ eigenvalues.
For the second five-brane:
\begin{equation}
M^{(2)} = \prod\gamma^{4}\gamma^{5}\gamma^{8}\gamma^{9}\gamma^{11}
\end{equation}
which has again two $+1$ and two $-1$ eigenvalues but in different
planes.
The surviving unbroken supersymmetry is given by the common eignevalues
of the two M matrices which gives a $N=2$ supersymmetry in the common $3
+ 1$ world-volume. The relevant chiral generators are
$$ Q_{1}^{\alpha} + {\tilde Q}^{\alpha}_{1};~~~{\rm and}~~~
Q_{2}^{\alpha} - {\tilde Q}^{\alpha}_{2}.
$$
The operator $Q_{i}^{\alpha}$ is given in the general form
$$
Q^{\alpha}_{i} \equiv \int e^{-\frac{1}{2}\phi}S^{\alpha}\Sigma_{i}
$$
where $\phi$ is the bosonization of superghost. $S^{\alpha}$ and $\Sigma_i$ 
are the
spin fields of the 4-dim. space-time and the internal parts respectively 
and they can be given explicitly in the bosonized form as follows:
$$
S^{\pm} = e^{\pm \frac{i}{2}(\phi_1 +\phi_2)}
$$
where $\phi_1$ and $\phi_2$ are the bosonization of the two complex fermions
associated to 4-dim. space-time. Similarly bosonising the complex 
fermions associated to 4-5, 6-7 and 8-9 planes respectively via scalars
$\phi_3$, $\phi_4$ and $\phi_5$ respectively, $\Sigma_k$ for $k=1,2$ can be 
expressed as: 
$$
\Sigma_k = e^{i(\frac{3}{2}-k)(\phi_3 +\phi_4) + i\frac{1}{2}\phi_5}
$$
Note that the spin field part coming from the mixed ND and DN directions
$\phi_3$ and $\phi_4$ appear with the same sign as in a $T^4/Z_2$ 
orbifold model and, together with the part coming from the DD directions,
it has the structure of the internal spin field of the heterotic string
on $T^2\times T^4/Z_2$, having $N=2$ spacetime supersymmetry in 4
dimensions.
Following the notation of \cite{agnt}, we write down the graviton vertex
operator in the 0-ghost picture,
$$
V_{gr}(p_1^{\mp}) =\bigl( \partial Z^{\pm}_{2} + 
ip_{1}^{\mp}\psi_{1}^{\pm}\psi^{\pm}_{2}\bigr)
\bigl( \dbar {\tilde Z}^{\pm}_{2} + ip_{1}^{\mp}{\tilde \psi}_{1}^{\pm}
{\tilde \psi}^{\pm}_{2}\bigr)e^{ip_{1}^{\mp}Z^{\pm}_{1}}.
$$
The graviphoton is obtained by the action of 
$( Q_{1}^{\alpha} + {\tilde Q}^{\alpha}_{1})( Q_{2}^{\alpha} - {\tilde 
Q}^{\alpha}_{2})$ on the graviton operator given above.
This gives rise to two types of terms; one from the action of
$(Q_{1}^{\alpha} Q_{2}^{\alpha} - {\tilde Q}^{\alpha}_{1}
{\tilde Q}^{\alpha}_{2})$
and the other from the action of 
$({\tilde Q}_{1}^{\alpha} Q_{2}^{\alpha} -  Q^{\alpha}_{1}{\tilde 
Q}^{\alpha}_{2})$. The first term is of the NS-NS type: 
$$
[\bigl(e^{-\phi}\psi^{+}_{5}\bigr)
\bigl( \dbar {\tilde Z}^{+}_{2} + ip_{1}^{-}{\tilde \psi}_{1}^{+}
{\tilde \psi}^{+}_{2}\bigr)-
\bigl( \partial Z^{+}_{2} + ip_{1}^{-}\psi_{1}^{+}\psi^{+}_{2}\bigr)
\bigl(e^{-{\tilde\phi}}{\tilde\psi}^{+}_{5}\bigr)]
e^{ip_{1}^{-}Z^{+}_{1}}.
$$
while the second one is of the R-R type:
$$
ip_{1}^{-}e^{-{1/2}({\phi + {\tilde \phi}})}S^{+}{\tilde
S}^{+}\Sigma_i {\tilde \Sigma}_j \epsilon_{ij}
$$
It is clear that the graviphoton is the sum of both these terms;
while the second is similar in form to the original type IIB graviphoton
operator ( compactified on Calabi-Yau), the first is similar in
structure to the gravi-photon in the heterotic string (compactified on
$T^{4}/Z_{2}\times  T^{2}$). Note also that the first term is actually the
anti-symmetric tensor component $B_{\mu I}$($I=8,9$) instead of the 
metric component $G_{\mu I}$. In fact the
latter is actually not gauge invariant: indeed one can check that its
longitudinal mode, although a total derivative, in general contributes to 
amplitudes involving world sheets with boundaries . More precisely the
boundary operator that it gives rise to is the scalar associated to the
overall translation of all the 5-branes. This is similar to what happens with
the NS $B_{\mu \nu}$ field whose longitudinal mode is the $U(1)$ gauge 
field partner of the above scalar. As a result these fields become 
massive due to Cremmer-Scherk mechanism.

\section{Computation of $F_g$}

The leading singularity in $F_g$ would appear in effective field theory 
as a one loop effect involving the would-be massless solitonic state 
propagating in the loop. Since this solitonic state is mapped to an
open string stretched between two intersecting $D$-branes the relevant
world-sheet is an annulus with the two boundaries on the two intersecting 
$D$-branes respectively. In order to calculate $F_g$ it turns out to be more 
convenient to
consider amplitudes involving $2g$ graviphoton field strengths that
appear as the lowest component of $W^{2g}$ and two matter $U(1)$ gauge
field strengths appearing in the highest component from the expansion of 
$F_g$. Since the leading singularity in $F_g$ is expected to be a constant 
times $\mu^{2-2g}$, where $\mu$ is the mass of the would-be massless soliton,
the relevant $U(1)$ gauge field strengths are the ones in the vector 
multiplet corresponding to the modulus $\mu$. This amplitude therefore 
will compute $\partial_{\mu}^2 F_g$. As mentioned earlier the vertex for 
this $U(1)$ gauge field is just the difference of the boundary operator: 
\begin{equation}
V_F = \int (\partial_{\tau} X^{\mu} +ip\psi \gamma_{\tau}\psi^{\mu})e^{ip.X}
\end{equation}
on the two boundaries of the annulus under consideration. 
We shall take a kinematic configuration so that $g$ of the graviphotons
are of the form:
\begin{equation}
V_T(p_1^{-})=\frac{1}{p_1^-}(Q_1^- +{\tilde Q}_1^-)(Q_2^- -{\tilde Q}_2^-)
V_{gr}(p_1^-)
\end{equation}
and similarly $g$ graviphotons with momenta $p_1^+$ which are obtained by 
exchanging the signs in the superscripts. 

These graviphoton vertex 
operators are in $(-1)$ ghost picture and therefore one needs to insert 
$2g$ picture changing operators on the annulus. Since each of the 
graviphoton vertex carries charge $1$ (left plus right charge) along
the DD plane, the only non-vanishing contribution comes from the part
($e^{\phi} e^{-\phi_5} \partial Z_5^{+} + {\rm right~ moving~ part}$) of the 
picture changing operator. Here $Z_5^+$ is the complex scalar associated to 
the DD plane. $\partial Z_5^+$ (and $\bar{\partial} Z_5^+$) are necessarily 
replaced by zero modes as they cannot contract among themselves. The zero 
mode contribution can be most easily seen when one represents the annulus 
by a rectangle defined by $0\leq \sigma \leq 1$ and $0 \leq \tau \leq t$,
where $t$ is the modulus of the world sheet. The various bosonic fields have
periodic boundary conditions along $\tau$ direction while along $\sigma$
direction they have the appropriate Neumann or Dirichlet boundary 
conditions. $Z_5^+$ then has the following mode expansion:
$$
Z_5^+ = \bar{\mu} \sigma + {\rm oscillators}
$$
where $\bar{\mu} = v_8 + i v_9$. Thus each picture changing operator 
gives a factor of $\bar{\mu}$ so that altogether we get a factor
$\bar{\mu}^{2g}$.

Now we construct the correlation functions between all the vertex 
operators. First thing to note is that in each of the operators appearing 
in the correlation function the ratio of the superghost charge and the 
charge along the DD plane (i.e. the coefficients of $\phi$ and $\phi_5$)
is $-1$. This, together with the fact that the spin-structures of the 
superghost
and the DD fermions are the same, implies that the correlation function
of the superghost system cancels exactly with that of DD fermion system.  
Finally we are left with the correlation functions of the remaining 4
complex fermion systems whose charges span a 4-dimensional lattice that
is closely related to $SO(8)$ weight lattices. More precisely since two of
the complex fermions come with mixed boundary conditions and therefore 
behave like $Z_2$ twisted fermions the $SO(8)$ weight lattices are shifted
by a spinor of the $SO(4)$. In any case, summing over the spin structures 
using Riemann theta identity is equivalent to using the triality relations
of the underlying $SO(8)$ weight lattices. If we denote by $a_i$ the 
charges of $\phi_i$ (for i=1,..,4) then one can show that summing over 
spin-structure is equivalent to the following triality map:
$$ a_1 \rightarrow (a_1 + a_2 + a_3 +a_4)/2 $$
$$ a_2 \rightarrow (a_1 + a_2 - a_3 -a_4)/2 $$
$$ a_3 \rightarrow (a_1 - a_2 + a_3 -a_4)/2 $$
$$ a_4 \rightarrow (-a_1 + a_2 + a_3 -a_4)/2 $$
The result of the spin structure therefore is correlation function of the 
transformed operators in the odd spin structure. Since the charges of all 
the operators appearing in our amplitude satisfy $a_1=a_2$ and $a_3=a_4$,
we note that the transformed operators do not carry any $\phi_3$ and 
$\phi_4$ charges. Thus the contribution of the $\phi_3$ and $\phi_4$ 
system exactly cancels the partition function of the bosons along DN and 
ND directions. Furthermore the bosonic partition function of the DD plane 
cancels that of the bosonic ghost system and we are just left with the 
correlation function of the space-time bosons and fermions. The above 
triality map results in the following transformations of the various 
operators appearing in the amplitude:
$$ Q_1^{+} \rightarrow \int dz\psi_1^{+}$$ 
$$ Q_1^{-} \rightarrow \int dz\psi_2^-$$
$$ Q_2^+ \rightarrow  \int dz\psi_2^+$$
$$ Q_2^- \rightarrow \int dz\psi_1^-$$
and similarly for the right-movers with $dz$ replaced by $d\bar{z}$, upto 
possible relative signs which can be absorbed in the definitions 
of $\psi$ and $\tilde{\psi}$'s.
$V_{gr}$ and $V_F$ under the triality map go to themselves. Note that the
supersymmetry generators involve the combinations 
$Q_1^{\pm}+\tilde{Q}_1^{\pm}$ and $Q_2^{\pm}-\tilde{Q}_2^{\pm}$. Under 
the triality map therefore they go over to the combinations 
$(\psi_1^{\pm}+  
\tilde{\psi}_1^{\pm})$ and $(\psi_2^{\pm}+ \tilde{\psi}_2^{\pm})$ 
respectively, where we have made a definite choice of relative signs 
between the left and the right movers. This choice together with the 
requirement of 
the closure of the contours involved in the supersymmetry charges then implies 
the following boundary conditions for $\psi_1^{\pm}$ and $\psi_2^{\pm}$:
$$\tilde{\psi}_1^{\pm}(z,\bar{z}) = \psi_1^{\pm}(-\bar{z},-z)$$
$$\tilde{\psi}_2^{\pm}(z,\bar{z}) = \psi_2^{\pm}(-\bar{z},-z)$$
By applying the supersymmetry generators on the graviton vertices one finds
that the graviphoton operators are
transformed by triality map to:
$$ V_T(p_1^{\mp}) \rightarrow \bigl{(} \partial_{\tau} Z_2^{\pm} 
+ip_1^{\mp}(\psi_1^{\pm}-  
\tilde{\psi}_1^{\pm})(\psi_2^{\pm}- \tilde{\psi}_2^{\pm})\bigr{)} 
e^{ip_1^{\mp}Z_1^{\pm}} $$

The correlation function of the bosonic part of the above vertex operator 
$\partial_{\tau} Z_2^{\pm}$ gives a total derivative which upon partial 
integration brings down one power of momentum $p_1^{\mp}$ together with 
$Z_1^{\pm}$ in each of the graviphoton vertex. One can now take the zero 
momentum limit after extracting one momentum from each vertex and then 
$F_g$ is related to a coorelation function involving $2g$ of the operators 
$ \bigl{(} Z_1^{\mp}\partial_{\tau} Z_2^{\pm} 
+(\psi_1^{\pm}-  
\tilde{\psi}_1^{\pm})(\psi_2^{\pm}- \tilde{\psi}_2^{\pm})\bigr{)}$.
One can now  
write down a generating function for $F_g$'s following \cite{agnt} as:
$$G(\lambda) = \sum_g \frac{\lambda^{2g}}{(g!)^2}\partial_{\mu}^2 F_g$$
which amounts to perturbing the sigma model action for space-time
bosons and fermions by 
$\lambda\bar{\mu}$ times the integral of the above operators and 
computing the correlation function of the two matter $U(1)$ gauge fields in
the presence of this perturbation.

The important point to note here is that the perturbed sigma model still 
admits four fermion zero modes given by $\tilde{\psi}_1^{\pm}= \psi_1^{\pm}
=$ constant and $\tilde{\psi}_2^{\pm}= \psi_2^{\pm}
=$ constant. Thus the partition function vanishes. This is to be expected
because the partition function computes an amplitude involving $2g$ 
graviphotons which is the lowest component of the corresponding $F$-term.
We would like to emphasize here that for this vanishing of the
partition function it is crucial that the full graviphoton vertex is used
in the computation including both the NS-NS part as well as R-R part.
In order to obtain a non-zero amplitude one must insert two $U(1)$
gauge fields as explained earlier. The two $U(1)$ gauge fields in our 
case are the vector partners of the modulus field $\mu$ and are given as
boundary operators. The four fermion zero modes are then soaked by the 
fermion bilinear pieces in the two vertices. Thus the generating function for
the $F_g$'s (or more precisely $\partial_{\mu}^2 F_g$'s) is obtained by 
computing the non-zero mode determinants of bosons and fermion in the 
perturbed action.

To compute the non-zero mode determinants let us make the mode expansion for
the non-zero mode parts of fermions and bosons:
$$ \psi_k^{\pm}= \sum_{(m,n)\neq(0,0)}\eta_k^{\pm 
(m,n)}e^{in\pi\sigma}e^{im\frac{2\pi}{t}\tau} $$
$$\tilde{\psi}_k^{\pm} = \sum_{(m,n)\neq(0,0)}\eta_k^{\pm 
(m,n)}e^{-in\pi\sigma}e^{im\frac{2\pi}{t}\tau} $$
$$ Z_k^{\pm}= \sum_{(m,n)\neq(0,0)} \alpha_k^{\pm(m,n)} cos (n\pi\sigma)
e^{im\frac{2\pi}{t}\tau}
$$
By plugging in these expansions one can easily see that for $m\neq 0$, the
boson and fermion determinants cancel. On the other hand for $m=0$, the 
$\lambda$-dependent term in the fermion part of the action drops out 
while it remains in the bosonic part of the action. The ratio of the 
fermion and boson determinant then gives
$$ \prod_{n=1}^{\infty}[1-(\frac{\lambda \bar{\mu}t}{n\pi})^2]^{-2} = 
[\frac{ \lambda \bar{\mu}t}{sin(\lambda \bar{\mu}t)}]^2
$$
Putting together the zero mode part of the bosons $Z_k^{\pm}$ as well as 
the boson $Z_5^{\pm}$ along DD directions one finds the following result for
the generating function:
\begin{eqnarray}
G(\lambda) \equiv \sum_{g=1}^{\infty}\lambda^{2g}\partial_{\mu}^2 F_g
&=& \int_{0}^{\infty} \frac{dt}{t} [\frac{
\lambda \bar{\mu} t}{sin(\lambda \bar{\mu}t)}]^2 e^{-|\mu|^2 t}
\nonumber \\ &=& \sum_{g=1}^{\infty} \frac{(2g-1)}{2g} B_{2g}  
\mu^{-2g}
\end{eqnarray}
where $B_{2g}$ are the Bernoulli numbers. Integrating this expression with 
respect to $\mu$, one finds the leading singularities in $F_g$ for $g\geq 2$:
$$ F_g \rightarrow \chi_g \mu^{2-2g}$$
where $\chi_g$ is the Euler character of the moduli space of genus $g$ 
Riemann surfaces. Similarly by integrating the equation for 
$G(\lambda)$ 
for $g=1$ we find: 
$$ F_1 \rightarrow -\frac{1}{12} \log \mu $$
The behaviour of the prepotential $F_0$ near the conifold can be easily 
calculated by inserting two $U(1)$ gauge fields corresponding to the 
modulus $\mu$ and the modulus field $\mu$ itself (the latter is inserted  
just to get non-vanishing on-shell amplitude). One can then verify that 
the prepotential has the expected behaviour near the conifold:
$$ F_0 \rightarrow \frac{1}{2} \mu^2 \log \mu $$

The above results are in agreement with the singularity structure one 
expects for $F_g$'s
near the conifold as due to the appearance of a massless hypermultiplet. Our
calculations here show that the description of the physics near conifold by
intersecting D-branes correctly reproduces the singularity structue. 
Note that in the computation for $F_g$'s all non-trivial dependence on the
world-sheet modulus $t$ drops out. This is to be expected, because only 
BPS-states contribute to $F_g$'s \cite{hm} and the 
only BPS state running through the loop in the above calculation is the 
ground state of the string stretched between the two intersecting D-branes.
Our computation also identifies the graviphoton vertex as a closed string 
operator involving both NS-NS and R-R parts. On the other hand the sum over 
all the $U(1)$'s associated to each $D$-brane would not reproduce this 
singularity structure as can be seen by using the method of \cite{bp}, and 
therefore cannot be identified with the graviphoton. In fact this is 
consistent with the observation that this sum over all $U(1)$'s is eaten 
up by the NS-NS $B_{\mu\nu}$ via Cremmer-Scherk mechanism.

{\bf{Acknowledgements}}

We would like to thank K. Varghese John, F. Morales, M. Serone and C. Vafa
for useful discussions. One of us (M.H.S.) acknowledges support by the
EEC contract SC1-CT92-0792.

\end{document}